Marcello Romano

# Exact Analytic Solutions for the Rotation of an Axially Symmetric Rigid Body Subjected to a Constant Torque

**Abstract** New exact analytic solutions are introduced for the rotational motion of a rigid body having two equal principal moments of inertia and subjected to an external torque which is constant in magnitude. In particular, the solutions are obtained for the following cases: (1) Torque parallel to the symmetry axis and arbitrary initial angular velocity; (2) Torque perpendicular to the symmetry axis and such that the torque is rotating at a constant rate about the symmetry axis, and arbitrary initial angular velocity; (3) Torque and initial angular velocity perpendicular to the symmetry axis, with the torque being fixed with the body. In addition to the solutions for these three forced cases, an original solution is introduced for the case of torque-free motion, which is simpler than the classical solution as regards

The author is with the
Mechanical and Astronautical Engineering Department, and with the Space Systems Academic Group, Naval Postgraduate School, 700 Dyer Road, Monterey, California
Tel.: ++1-831-656-2885
Fax: ++1-831-656-2238
E-mail: mromano@nps.edu

its derivation and uses the rotation matrix in order to describe the body orientation. This paper builds upon the recently discovered exact solution for the motion of a rigid body with a spherical ellipsoid of inertia. In particular, by following Hestenes' theory, the rotational motion of an axially symmetric rigid body is seen at any instant in time as the combination of the motion of a "virtual" spherical body with respect to the inertial frame and the motion of the axially symmetric body with respect to this "virtual" body. The kinematic solutions are presented in terms of the rotation matrix. The newly found exact analytic solutions are valid for any motion time length and rotation amplitude. The present paper adds further elements to the small set of special cases for which an exact solution of the rotational motion of a rigid body exists.

## 1 Introduction

The study of angular motion of a rigid body is at the very basis of classical physics. Besides the theoretical interest, the subject is of considerable practical importance for the fields of astronautics and celestial mechanics.

The problem of the rotational motion of a rigid body can be divided into two parts. The *dynamic problem* aims to obtain the angular velocity of the body with respect to an inertial reference by starting from the knowledge of the initial angular velocity and the history of the applied torque. On the other hand, the *kinematic problem* focuses on determining the current orientation of the body from the knowledge of the initial orientation and the history of the angular velocity.

The exact analytic solution for the rotational motion of a rigid body exists only for a small number of special cases.

When no external torque (other than the one due to gravity) acts on the body, exact solutions exist for both the dynamic and kinematic problems only in the following three cases:

1. Euler-Poinsot case: torque-free rotation of an asymmetric rigid body with a fixed point coincident with the center of mass (Euler 1758; Poinsot 1834; Jacobi 1849; Leimanis 1965; Morton et al. 1974);
2. Lagrange-Poisson heavy top case: rigid body under gravity force with two equal principal moments of inertia at the fixed point and the center of mass along the third axis of inertia (Lagrange 1788; Poisson 1813; Golubev 1960; Leimanis 1965);
3. Kovalevskaya heavy top case: rigid body under gravity force with two equal principal moments of inertia at the fixed point, value of the third moment of inertia equal to half of the value of the moment inertia about the other two axes, and the center of mass in the plane of equal moments of inertia (Kovalevskaya 1890; Leimanis 1965).

On the contrary, when the rigid body is subjected to an external torque, the complete (i.e. for both the kinematics and the dynamics) exact analytic solution exists only for a rigid body with spherical ellipsoid of inertia subjected to a constant torque and arbitrary initial angular velocity (Romano 2008), in addition to the straightforward single degree-of-freedom case of rotation with a torque along a principal axis of inertia and initial angular velocity along the same axis.

Furthermore, a partial (limited to the dynamic problem) exact analytic solution exists for the case of an axially symmetric body subjected to a constant torque (Longuski 1991, Tsiotras and Longuski 1991). In particular, Longuski 1991 and Tsiotras and Longuski 1991 use a complex form expression of the Euler's dynamic equations and give an exact solution for the angular velocity involving the Fresnel integral function, and an approximate solution for the kinematic problem in terms of Euler angles. Nevertheless, no exact analytic solution has been found for the kinematic problem.

Several researchers have proposed approximate solutions. In particular, Longuski 1991, and Longuski and Tsiotras 1995 use asymptotic and series expansion to get the approximate solution of the dynamics and kinematics for both the constant torque case and the more difficult case of time-varying torque. Livneh and Wie 1997 analyze qualitatively, in the phase space, the dynamics of a rigid-body subjected to constant torque. As regards the kinematic problem, Iserles and Nørsett 1999 study the solution in terms of series expansion for the more general problem of solving linear differential equation in Lie Groups, building upon the work of Magnus 1954. Finally, Celledoni and Saefstroem 2006 propose ad-hoc numerical integration algorithms.

For the first time, to the knowledge of the author, the present paper introduces the complete exact analytic solutions for the rotation of an axially symmetric rigid body in the following three cases

1. External torque constant in magnitude and parallel to the symmetry axis, with arbitrary initial angular velocity and body orientation.
2. External torque constant in magnitude, perpendicular to the symmetry axis and rotating at a constant rate about the symmetry axis, with arbitrary initial angular velocity and body orientation.
3. External torque and initial angular velocity perpendicular to the symmetry axis, with the torque constant in magnitude and fixed with the body and arbitrary initial body orientation.

The solution for the kinematics are given in terms of the rotation matrix.

This paper builds upon the recently found exact analytic solution for the motion of a rigid body with a spherical ellipsoid of inertia (Romano 2008), and the Hestenes' method of decomposing the rotational motion of an axially symmetric rigid body into the combination of the motion of a "virtual" spherical body with respect to the inertial frame and the motion of the axially symmetric body with respect to this "virtual" body (adapted from Hestenes 1999).

In addition to the three cases listed above, an original complete solution of the Euler-Poinsot's classical case of torque-free motion is also presented in this paper, which is simpler than the classical solution (Leimanis 1965) as regards its derivation and uses the rotation matrix in order to describe the body orientation.

The newly introduced exact analytic solutions are of high theoretical interest, as they add new elements to the set of few special cases for which an exact solution of the rotational motion of a rigid body exists. Additionally, they are also of interest from the applied mathematics point of view as they constitute significant new comparison cases for the validation and error analysis of approximate algorithms. Indeed, the analytic solutions presented in this paper are valid for any length of time and rotation amplitude: they only require the numerical evaluation of mathematical expressions, without needing any numerical propagation.

Finally, the solutions introduced in this paper are valid for any rigid body which has an ellipsoid of inertia of revolution, i.e. two equal principal moments of inertia. This class of bodies, having a *dynamic* axial symmetry, contains as a subset the class of bodies having *geometric* axial symmetry (Hestenes 1999). However, following the common practice adopted in literature, in this paper the definition of "axially symmetric body" and "body with revolution ellipsoid of inertia" are used interchangeably.

The paper is organized as follows: Section 2 briefly introduces the dynamic and kinematic equations for the rotational motion of a generic rigid body and Section 3 summarizes the known results for a rigid body having spherical ellipsoid of inertia. Section 4 introduces the method used to analyze the motion of a rigid body with revolution ellipsoid of inertia, and Section 5 presents the new exact analytic solutions listed above. Finally, Section 6 concludes the paper.

## 2 General Rigid-Body Problem Statement

For a generic rigid body the Euler's rotational equation of motion, in vectorial form, is (Goldstein 1980)

$$\dot{\underline{h}} = \underline{m} \tag{1}$$

where the the dot symbol denotes the time derivative with respect to an inertial reference frame, $\underline{m}$ is the resultant external torque acting on the body, and

$$\underline{h} = \underline{\mathcal{I}}\,\underline{\omega} \tag{2}$$

is the absolute angular momentum of the body, $\underline{\mathcal{I}}$ is the inertia dyadic of the body with respect to its center of mass, and $\underline{\omega}$ is the angular velocity of the body with respect to an inertial reference frame.

By resolving all of the vectors and the inertia dyadic along a body-fixed Cartesian coordinate system $B$ with axis equal to the principal axes of inertia, Eq. 1 can be written in scalar form as

$$\begin{aligned} I_1 \dot{p} &= (I_2 - I_3)\,q\,r + m_1 \\ I_2 \dot{q} &= (I_3 - I_1)\,r\,p + m_2 \\ I_3 \dot{r} &= (I_1 - I_2)\,p\,q + m_3, \end{aligned} \tag{3}$$

where $\{I_i : i = 1, 2, 3\}$ are the principal moments of inertia, $\{p, q, r\}$ are the components in the coordinate system $B$ of $\underline{\omega}$, and elements of the column matrix $^B\omega_{BN}$. Finally $\{m_i : i = 1, 2, 3\}$ are the components of $\underline{m}$ along the same axes.

The rotation matrix $R_{NB} \in SO(3)$ from the body fixed coordinate system $B$ to an inertial coordinate system $N$ obeys the following differential equation (Leimanis 1965)

$$\dot{R}_{NB} = R_{NB}\,\Omega\left(^B\omega_{BN}\right), \tag{4}$$

where

$$\Omega\left({}^{B}\omega_{BN}\right) = \begin{bmatrix} 0 & -r & q \\ r & 0 & -p \\ -q & p & 0 \end{bmatrix}, \qquad (5)$$

In general, Equations 3 and 4, which solve the dynamic and kinematic motion problem of a rigid body respectively, do not have exact analytic solutions. In particular, the following straightforward direct solution of Eq. 4

$$R_{NB}(t) = R_{NB}(0) \, \exp\left(\int_0^t \Omega(\xi)\, d\xi\right), \qquad (6)$$

where $\exp(\bullet)$ indicates the matrix exponential, is only valid for the very particular cases when the matrix $\Omega(t)$ commutes with its integral (Adrianova 1995). This commutativity is lost when the initial angular velocity is arbitrary; therefore, Eq. 6 is not valid anymore in that case. The limitation in the validity of the solution of Eq. 6 is not clearly stated in Bödewadt 1952 and Leimanis 1965, as already noticed in Longuski 1984.

**3 Exact analytic solution for the motion of a rigid body with a spherical ellipsoid of inertia**

This section summarizes the results introduced by Romano 2008 for the motion of a rigid body with spherical ellipsoid of inertia and subjected to a constant torque. The developments reported in this section will be used in the following sections of this paper in order to analyze the motion of an axially symmetric rigid body.

For a rigid body having a spherical ellipsoid of inertia, the kinematic differential equations in terms of the stereographic complex rotation variables $w_k$ (Tsiotras and Longuski 1991, Schaub and Junkins 2003) are (Romano 2008)

$$\dot{w}_k = \frac{1}{2}\,(p_0 - i\,q_0)\,w_k^2 - i\,(r_0 + U\,t)\,w_k + \frac{1}{2}\,(p_0 + i\,q_0), \quad k = 1, 2, 3. \quad (7)$$

**Theorem 1** *Given $p_0, q_0, r_0$ and $U$ real numbers, with $p_0$ and $q_0$ not both zero, the general solution for each one of the Eq. 7, governing the rotational kinematics of a rigid body with spherical ellipsoid of inertia, initial angular velocity components $p(0) = p_0$, $q(0) = q_0$, and $r(0) = r_0$ along the three body fixed axes $\sigma_1, \sigma_2$ and $\sigma_3$, and subjected to a constant torque $U$, normalized by the value of the moment of inertia and directed along the axis $\sigma_3$, is the following*

$$w(t,c) = \frac{(1+i)\sqrt{U}}{3(p_0 - iq_0)} \left[6z + G(z,c)\right], \tag{8}$$

with

$$G(z,c) := \frac{2\,_1F_1\left(\frac{3-\nu}{2}, \frac{5}{2}; z^2\right)(\nu-1)z^2 + 6\,_1F_1\left(1 - \frac{\nu}{2}, \frac{3}{2}; z^2\right)c\nu z - 3\,_1F_1\left(\frac{1-\nu}{2}, \frac{3}{2}; z^2\right)}{_1F_1\left(-\frac{\nu}{2}, \frac{1}{2}; z^2\right)c + \,_1F_1\left(\frac{1-\nu}{2}, \frac{3}{2}; z^2\right)z}, \tag{9}$$

*where $_1F_1$ denotes the confluent hypergeometric function (Lebedev 1965), $c \in \mathbb{C}$ is the constant of integration and*

$$z := \frac{(1+i)(r_0 + Ut)}{2\sqrt{U}}, \qquad \nu := -1 - \frac{i(p_0^2 + q_0^2)}{4U}. \tag{10}$$

**Corollary 1** *The solution in terms of the rotation matrix, which corresponds to the solution given by theorem 1 in terms of stereographic rotation variables, is*

$$R_{NB}(t) = [r_{kj}(t)], \quad k, j = 1, 2, 3, \tag{11}$$

with

$$r_{k1} = \frac{i(w_k - \overline{w_k})}{1 + |w_k|^2}, \quad r_{k2} = \frac{w_k + \overline{w_k}}{1 + |w_k|^2}, \quad r_{k3} = \frac{1 - |w_k|^2}{1 + |w_k|^2}, \quad k = 1, 2, 3, \tag{12}$$

*where $w_k = w(t, c_k)$, being $w(t, c_k)$ given by Eq. 8 with the following values for the initial conditions (obtained by considering the rotation matrix $R_{NB}$ to be equal to the identity matrix at the initial time $t = 0$, without loss of generality).*

$$c_1 = -\frac{(1+i)}{6\sqrt{U}} \left\{ \frac{_1F_1\left(\frac{1-\nu}{2}, \frac{3}{2}; \frac{ir_0^2}{2U}\right)\left[6r_0^2 + 3(p_0 - iq_0)r_0 + 6iU\right] + 2g}{2\,_1F_1\left(1 - \frac{\nu}{2}, \frac{3}{2}; \frac{ir_0^2}{2U}\right)\nu r_0 + \,_1F_1\left(-\frac{\nu}{2}, \frac{1}{2}; \frac{ir_0^2}{2U}\right)(p_0 - iq_0 + 2r_0)} \right\}$$

$$c_2 = \frac{(1-i)}{6\sqrt{U}} \left\{ \frac{{}_1F_1\left(\frac{1-\nu}{2}, \frac{3}{2}; \frac{ir_0^2}{2U}\right)\left[6ir_0^2 - 3(p_0 - iq_0)r_0 - 6U\right] - 2g}{2\,{}_1F_1\left(1 - \frac{\nu}{2}, \frac{3}{2}; \frac{ir_0^2}{2U}\right)\nu r_0 + {}_1F_1\left(-\frac{\nu}{2}, \frac{1}{2}; \frac{ir_0^2}{2U}\right)(ip_0 + q_0 + 2r_0)} \right\}$$

$$c_3 = \frac{(1+i)}{6r_0\sqrt{U}} \left\{ \frac{g - 3\,{}_1F_1\left(\frac{1-\nu}{2}, \frac{3}{2}; \frac{ir_0^2}{2U}\right)(r_0^2 + iU)}{{}_1F_1\left(1 - \frac{\nu}{2}, \frac{3}{2}; \frac{ir_0^2}{2U}\right)\nu + {}_1F_1\left(-\frac{\nu}{2}, \frac{1}{2}; \frac{ir_0^2}{2U}\right)} \right\}, \qquad (13)$$

where $\nu$ is defined as in Eq. 10, and

$$g := {}_1F_1\left(\frac{3-\nu}{2}, \frac{5}{2}; \frac{ir_0^2}{2U}\right)(\nu - 1)r_0^2. \qquad (14)$$

In the general case when the constant torque is not directed along the axis $\sigma_3$, and the coordinate system $B$ is not coincident with $N$ at the initial time, the results of Theorem 1 and Corollary 1 are still applicable, as reported in the following corollary, which constitutes a new development with respect to Romano 2008.

**Corollary 2** *Let us consider that the direction of the external torque is identified in the coordinate system $B$ by the unit vector*

$$u = \left\{ \begin{array}{c} u_1 \\ u_2 \\ u_3 \end{array} \right\}, \qquad (15)$$

*and that the initial orientation of $B$ with respect to the inertial coordinate system $N$ is given by $R_{BN}(0)$. Then, by exploiting the property of successive rotations, it results*

$$R_{BN}(t) = R_{BK}\,R_{KN}(t)\,R_{KB}\,R_{BN}(0), \qquad (16)$$

*where $R_{KB}$ is any time-independent rotation matrix which brings the direction of the torque to coincide with the third axis of the auxiliary body-fixed coordinate system $K$. In other words, the matrix $R_{KB}$ needs to satisfy the condition*

$$R_{KB} \left\{ \begin{array}{c} u_1 \\ u_2 \\ u_3 \end{array} \right\} = \left\{ \begin{array}{c} 0 \\ 0 \\ 1 \end{array} \right\}. \qquad (17)$$

For instance, $R_{KB}$ can be obtained by combining a first elementary rotation of an angle $\alpha = \arctan 2(u_2, u_1)$ (indicating by $\arctan 2$ the four quadrant inverse tangent) about the third axis of $B$ with a second elementary rotation of an angle $\beta = \arctan 2(u_2/sin(\alpha), u_3)$ about the resulting second axis, yielding

$$R_{KB} = \begin{bmatrix} \cos(\beta) & 0 & -\sin(\beta) \\ 0 & 1 & 0 \\ \sin(\beta) & 0 & \cos(\beta) \end{bmatrix} \begin{bmatrix} \cos(\alpha) & \sin(\alpha) & 0 \\ -\sin(\alpha) & \cos(\alpha) & 0 \\ 0 & 0 & 1 \end{bmatrix}. \quad (18)$$

Furthermore, $R_{KN}(t)$ is obtained by transposing the resulting matrix of Eq. 11 of Corollary 1, with the subindex $B$ substituted by $K$, and the following values of the components along the coordinate system $K$ of the initial angular velocity of the body with respect to the inertial frame

$$\begin{Bmatrix} p_0 \\ q_0 \\ r_0 \end{Bmatrix} = R_{KB} \begin{Bmatrix} p'_0 \\ q'_0 \\ r'_0 \end{Bmatrix}, \quad (19)$$

where the prime symbol is used to denote the initial conditions expressed in the $B$ coordinate system.

## 4 Analysis of the rotational motion of a rigid body with a revolution ellipsoid of inertia

Let us assume, from now on, that the rigid body has ellipsoid of inertia which is a revolution ellipsoid. Consequently, let us assume that the principal moments of inertia satisfy the following relation

$$I_1 = I_2 = I \neq I_3. \quad (20)$$

By following the development of Hestenes 1999, the absolute angular momentum of the body can be expressed by

$$\underline{h} = \underline{\mathcal{I}}\,\underline{\omega} = I\underline{\omega} + (I_3 - I)(\underline{\omega} \cdot \underline{e})\,\underline{e}, \quad (21)$$

where $\underline{e}$ is the unit vector along the direction of the symmetry axis.

Additionally, the angular velocity can be seen, at any instant of time, as the sum of two vectorial components, whose one is parallel to the angular momentum vector ($\underline{\omega}_h$) and the other one is parallel to the body symmetry axis ($\underline{\omega}_e$), i.e.

$$\underline{\omega} = \underline{\omega}_h + \underline{\omega}_e. \tag{22}$$

In particular, from Eq. 21, it yields

$$\underline{\omega}_h = \frac{\underline{h}}{I}, \ \underline{\omega}_e = \frac{(I - I_3)}{I_3} \left( \frac{\underline{h}}{I} \cdot \underline{e} \right) \underline{e} = A(\underline{\omega}_h \cdot \underline{e}) \underline{e}, \tag{23}$$

with the constant $A$ defined as $A = (I - I_3)/I_3$.

From Eq. 23 it results

$$\underline{h} = I \underline{\omega}_h, \tag{24}$$

and, from the Euler's equation (Eq. 1),

$$\underline{\dot{h}} = I \underline{\dot{\omega}}_h = \underline{m} \tag{25}$$

Equations 22 and 25 are the mathematical expression of the *Reduction Theorem* (Hestenes 1999) which can be formally stated as follows: the evolution in time of the absolute angular momentum of an axially symmetric body subjected to an external torque vector $\underline{m}$ is corresponding to the motion of a "virtual" homogeneous spherical body with the same value of the transversal inertia of the axially symmetric body and subjected to the same external torque.

In other words, to exemplify, the rotational motion of an axially symmetric rigid body (with two equal moments of inertia $I$ and third moment of inertia $I_3$) is decomposed into the combination of the motion of a "virtual" spherical body (with principal moment of inertia $I$) with respect to the inertial frame and the spinning motion, about its symmetry axis, of the axially symmetric rigid body with respect to this "virtual" spherical body.

In order to analyze in more details the motion of the axially symmetric body, we consider the following three Cartesian coordinate systems:

1. A principal coordinate system $B$ attached to the axially symmetric body and centered at its center of mass.
2. A coordinate system $S$ attached to the "virtual" spherical body and centered at its center of mass.
3. An inertially fixed coordinate system $N$.

*Assumptions 1.* Without loosing generality we assume that the coordinate system $B$ has its third axis parallel to the axis of symmetry of the body, and that the coordinate systems $B$ and $S$ have superimposed axes at the initial time ($t = 0$), i.e. that $R_{BS}(0)$ is an identity matrix.

We now take advantage of the reduction theorem by subdividing the procedure to reach the final goal of solving the motion of the coordinate system $B$ with respect to the inertial frame into two subproblems: first we solve for the motion of the coordinate system $B$ with respect to $S$, and then we look for a solution of the motion of the coordinate system $S$ with respect to $N$. The advantage of this procedure stays in the fact that the problem of solving the motion for an axially symmetric body is reduced to the simpler problem of solving the motion for a body having spherical ellipsoid of inertia.

Because of the definitions of $\underline{\omega}_h$ and $\underline{\omega}_e$ (see Eq 22 and Eq. 23) and of the Assumptions 1, the coordinate system $B$ is moving with respect to the coordinate system $S$ by rotating about its third axis with the angular rate

$$\omega_e = |\underline{\omega}_e| = A(\underline{\omega}_h \cdot \underline{e}). \tag{26}$$

Indeed, the vector $\underline{\omega}_h$ is the angular velocity of the coordinate system $S$ with respect to the coordinate system $N$.

In particular, the solution for the kinematic description of the motion of $B$ with respect to $S$ can be stated as follows (see also Eq. 4 and 6)

$$R_{BS}(t) = R_{BS}(0) \exp\left(\int_0^t \Omega\left(^S\omega_{SB}\right)(\xi)\, d\xi\right) = \begin{bmatrix} \cos[f(t)] & \sin[f(t)] & 0 \\ -\sin[f(t)] & \cos[f(t)] & 0 \\ 0 & 0 & 1 \end{bmatrix}, \tag{27}$$

with

$$f(t) = \int_0^t \omega_e(\xi)\, d\xi. \tag{28}$$

In developing Eq.27 it has been taken into account that $R_{BS}(0)$, without loss of generality, is an identity matrix because of the Assumptions 1, and

$$^S\omega_{SB} = {}^B\omega_{SB} = \begin{Bmatrix} 0 \\ 0 \\ -\omega_e(t) \end{Bmatrix}, \quad \Omega\left({}^S\omega_{SB}\right)(t) = \begin{bmatrix} 0 & \omega_e(t) & 0 \\ -\omega_e(t) & 0 & 0 \\ 0 & 0 & 0 \end{bmatrix}. \tag{29}$$

In order to fully solve for the motion of the axially symmetric body it now remains to find the solution for the motion of the coordinate system $S$ with respect to the coordinate system $N$.

In particular, the dynamics of the 'virtual" spherical body with respect to the inertial frame is governed by Eq. 25. By expressing all of the vectors of that equation in scalar components along the coordinate system $S$, it yields

$$^S\dot{\omega}_h = \frac{1}{I}{}^S m = \frac{1}{I} R_{SB}{}^B m, \tag{30}$$

where $^S\omega_h$ is the column matrix obtained by projecting the angular velocity vector $\underline{\omega}_h$ along the coordinate system $S$, defined as

$$^S\omega_h = {}^S\omega_{SN} = \begin{Bmatrix} \bar{p} \\ \bar{q} \\ \bar{r} \end{Bmatrix}, \tag{31}$$

and $^B m$ is the column matrix obtained by projecting the torque vector $\underline{m}$ along the coordinate system $B$.

The initial conditions for Eq. 30 are (see Eq. 24 and Assumptions 1)

$$^S\omega_{SN}(0) = \begin{Bmatrix} \bar{p}_0 \\ \bar{q}_0 \\ \bar{r}_0 \end{Bmatrix} = \frac{R_{SB}(0)\,{}^B h_0}{I} = \begin{Bmatrix} p_0 \\ q_0 \\ \frac{I_3}{I} r_0 \end{Bmatrix}, \tag{32}$$

where $p_0$, $q_0$ and $r_0$ are the initial conditions for the angular rate of the rigid body with revolution ellipsoid of inertia.

Furthermore, from Eq. 26 it yields

$$\omega_e(t) = A\,\bar{r}(t). \tag{33}$$

In particular, for all of the cases when the external acting torque depends at most on time, from Equations 30 it results

$$\omega_e(t) = A\left(\frac{I_3}{I}r_0 + \frac{1}{I}\int_0^t m_e(\xi)\,d\xi\right), \tag{34}$$

where $m_e = \underline{m}\cdot\underline{e}$.

The kinematic equation in terms of motion of the coordinate system $S$ with respect to $N$ is

$$\dot{R}_{NS} = R_{NS}\,\Omega\left(^S\omega_h\right). \tag{35}$$

Finally, once obtained the solution for the Equations 34, 27, 30 and 35, the complete solution for the dynamic description of the motion of the axially symmetric rigid body is given at each time instant $t$ by

$$^B\omega_{BN}(t) = {}^B\omega_{BS}(t) + R_{BS}(t)\,{}^S\omega_{SN}(t), \tag{36}$$

while the solution of the correspondent kinematic problem is

$$R_{BN}(t) = R_{BS}(t)\,R_{SN}(t). \tag{37}$$

Indeed, Eq. 36 is immediately obtained from Eq. 37 by taking the time derivative of both sides of Eq. 37, by considering Equations 4 and 5, and by exploiting the property

$$\Omega\left(R_{BS}\,{}^S\omega_{SN}\right) = R_{BS}\,\Omega\left(^S\omega_{SN}\right)R_{SB}. \tag{38}$$

**5 Exact analytic solutions for the motion of a rigid body with a revolution ellipsoid of inertia**

By leveraging the developments of previous sections, this section introduces the exact analytic solutions for the dynamics and kinematics of a rigid body having ellipsoid of inertia of revolution. Four cases are considered in the following four subsections.

5.1 Case with a constant external torque parallel to the symmetry axis

The following theorem introduces a previously unknown exact analytic solution for the complete dynamic and kinematic problems of an axially symmetric rigid body with a constant external torque staying at any time parallel to the symmetry axis, and arbitrary initial orientation and angular velocity.

**Theorem 2** *Given $p_0, q_0, r_0$ and $U_3$ real numbers, the solution of the dynamic problem of determining, at any time $t$, the absolute angular velocity of a rigid body having two equal principal moments of inertia about the principal body axes ($\sigma_1$) and ($\sigma_2$), initial angular velocity components $p(0)=p_0$, $q(0)=q_0$, and $r(0)=r_0$ along the three principal axes, and subjected to a constant torque $U_3$, normalized by the value of the equal principal moment of inertia ($I$) and directed along the unequal inertia principal axis ($\sigma_3$), is the following*

$$^B\omega_{BN}(t) = \begin{Bmatrix} p_0 \cos\left[f(t)\right] + q_0 \sin\left[f(t)\right] \\ -p_0 \sin\left[f(t)\right] + q_0 \cos\left[f(t)\right] \\ r_0 + \frac{I}{I_3} U_3 t \end{Bmatrix}, \qquad (39)$$

*while the solution of the correspondent kinematic problem of determining, at any time $t$, the orientation of the body with respect to the inertial frame, i.e. of the coordinate system $B$ with respect to $N$, is given by*

$$R_{BN}(t) = R_{BS}(t)\, R_{SN}(t) \qquad (40)$$

*where $R_{SN}(t)$ is obtained from Eq. 16 of Corollary 2, by substituting the subindex $B$ by $S$, the $R_{BK}$ matrix by the identity matrix, the values of the initial conditions by $p'_0 = p_0, q'_0 = q_0, r'_0 = \frac{I_3}{I} r_0$ and the value of the torque magnitude by $U = U_3$. Finally, $R_{BS}(t)$ is given by Eq. 27 with*

$$f(t) = A\left(\frac{I_3}{I} r_0 t + \frac{U_3}{2} t^2\right). \qquad (41)$$

*Proof* In this case we suppose to have the following external torque acting on the axially symmetric rigid body

$$^B m(t) = \begin{Bmatrix} 0 \\ 0 \\ m_3 \end{Bmatrix}, \qquad (42)$$

with $m_3$ a scalar constant.

From Equations 27 and 42, it results that the external torque has the same scalar components as seen from both the coordinate system $B$ and the coordinate system $S$, which is attached to the "virtual" spherical body; this results from the fact that the torque is directed along the rotation axis between the two coordinate systems. Therefore, it yields

$$^S m(t) = R_{SB}(t)\, ^B m(t) = {}^B m(t). \qquad (43)$$

By taking into account the Equations 42 and 43, the solution of the dynamic problem for the "virtual" spherical body, i.e. of Eq. 30 with initial conditions 32, is immediately given by

$$^S \omega_{SN}(t) = \begin{Bmatrix} \bar{p}(t) \\ \bar{q}(t) \\ \bar{r}(t) \end{Bmatrix} = \begin{Bmatrix} p_0 \\ q_0 \\ \frac{I_3}{I} r_0 + U_3 t \end{Bmatrix}, \qquad (44)$$

where $U_3 = m_3/I$.

The solution of the kinematic problem for the "virtual" spherical body correspondent to the kinematic solution of Eq.44, stated in Eq. 35 , i.e. the determination of the rotation matrix $R_{SN}(t)$, is given by the Corollary 2 (see Section 3 and the statement of Theorem 2).

Now, in order to reach the goal of fully solving the motion of the axially symmetric rigid body with respect to the inertial frame, it just remains to find the solution for the motion of the coordinate system $B$ with respect to the coordinate system $S$.

In particular, $^B\omega_{SB}$ is immediately given by Eq. 29 with (from Equations 34 and 44)

$$\omega_e(t) = A\left(\frac{I_3}{I} r_0 + U_3 t\right). \tag{45}$$

Furthermore, the motion of the coordinate system $B$ with respect to $S$ is governed by Eq. 27 with

$$f(t) = \int_0^t \omega_e(\xi)\, d\xi = A\left(C + \frac{I_3}{I} r_0\, t + \frac{U_3}{2} t^2\right), \tag{46}$$

where $C = 0$ because of Assumptions 1.

Finally, the complete solution of the dynamic problem is given by Eq. 39, which follows from the Equations 36, 44, and 45, while the solution of the kinematic problem is given by Eq. 40.

Theorem 2 results therefore proven.

5.2 Cases with a constant external torque perpendicular to the symmetry axis

This section introduces the exact analytic solution for the complete dynamic and kinematic problems of an axially symmetric rigid body with an external torque which is constant in magnitude and stays at any time perpendicular to the symmetry axis. Two cases are presented, in two different theorems.

In particular, the following theorem introduces a previously unknown exact analytic solution for an axially symmetric rigid body with a constant external torque staying at any time perpendicular to the symmetry axis and rotating at a constant rate about the symmetry axis, with arbitrary initial angular velocity and body orientation.

**Theorem 3** *Given $p_0, q_0, r_0$ and $U$ real numbers, the solution of the dynamic problem of determining, at any time $t$, the absolute angular velocity of a rigid body having two equal principal moments of inertia ($I$) about the principal body axes ($\sigma_1$) and ($\sigma_2$) and principal moment of inertia $I_3$ about the third axis ($\sigma_3$), initial angular velocity components $p(0) = p_0$, $q(0) = q_0$, and* $r(0) = r_0$ along the three principal axes, and subjected to a constant torque perpendicular to the third axis, rotating about the third axis at the angular rate $\left(\frac{I_3-I}{I}\right) r_0$, and having magnitude $U$ normalized by the value of the equal principal moment of inertia $(I)$, is the following

$$^B\omega_{BN}(t) = \begin{Bmatrix} (p_0 + U\,t) \cos[f(t)] + q_0 \sin[f(t)] \\ -(p_0 + U\,t) \sin[f(t)] + q_0 \cos[f(t)] \\ r_0 \end{Bmatrix}, \qquad (47)$$

while the solution of the correspondent kinematic problem of determining, at any time $t$, the orientation of the body with respect to the inertial frame, i.e. of the coordinate system $B$ with respect to $N$, is given by

$$R_{BN}(t) = R_{BS}(t)\,R_{SN}(t), \qquad (48)$$

where $R_{SN}(t)$ is obtained from Eq. 16 of Corollary 2, by substituting the subindex $B$ by $S$, the time-independent matrix $R_{BK}$ by

$$R_{SK} = \begin{bmatrix} 0 & 0 & 1 \\ 0 & 1 & 0 \\ -1 & 0 & 0 \end{bmatrix}, \qquad (49)$$

the values of the initial conditions by $p'_0 = p_0, q'_0 = q_0, r'_0 = \frac{I_3}{I} r_0$. Finally, $R_{BS}(t)$ is given by Eq. 27 with

$$f(t) = \left(\frac{I - I_3}{I}\right) r_0\,t. \qquad (50)$$

*Proof* In case of external torque perpendicular to the symmetry axis of the body, the dynamics and kinematics of the motion of the coordinate system $B$ with respect to the coordinate system $S$ are independent of the acting torque. Indeed, Eq. 34, by taking into account that $m_e = 0$, yields

$$\omega_e = A\left(\frac{I_3}{I} r_0\right) = \left(\frac{I - I_3}{I}\right) r_0. \qquad (51)$$

Then $^B\omega_{SB}$ is immediately obtained by inserting Eq. 51 into Eq. 29. Furthermore, $R_{BS}(t)$ is given by Eq. 27 with

$$f(t) = \omega_e\,t \qquad (52)$$

In order to reach the goal of fully solving the motion of the axially symmetric rigid body with respect to the inertial frame, it just remains to be solved the problem of finding the motion of the "virtual" spherical body with respect to the inertial frame, i.e. the motion of the coordinate system $S$ with respect to the coordinate system $N$.

Because of the assumption made in the statement of Theorem 3 that the external torque is rotating about the axis of symmetry at a constant angular rate equal to $-\omega_e$, it yields

$$^B m(t) = \begin{Bmatrix} m \cos(\omega_e t) \\ -m \sin(\omega_e t) \\ 0 \end{Bmatrix}, \tag{53}$$

with $m$ a scalar constant.

From Equations 27 and 53, it results that the external torque has the following expression in components along the $S$ coordinate system

$$^S m(t) = R_{SB}(t) \,^B m(t) = \begin{Bmatrix} m \\ 0 \\ 0 \end{Bmatrix}. \tag{54}$$

By taking into account the Equations 53 and 54, the solution of the dynamic problem for the "virtual" spherical body, i.e. of Eq. 30, is immediately given by

$$^S \omega_{SN}(t) = \begin{Bmatrix} \bar{p}(t) \\ \bar{q}(t) \\ \bar{r}(t) \end{Bmatrix} = \begin{Bmatrix} p_0 + Ut \\ q_0 \\ \frac{I_3}{I} r_0 \end{Bmatrix}, \tag{55}$$

where $U = m/I$.

The solution of the kinematic problem for the "virtual" spherical body correspondent to the kinematic solution of Eq. 55 is given by Corollary 2.

Finally, the complete solution of the dynamic problem is given by Eq. 47, as it follows from the Equations 36, 51, and 55, while the solution of the kinematic problem is given by Eq. 48.

Theorem 3 results therefore proven.

The following theorem introduces the exact analytic solution for the complete dynamic and kinematic problems of an axially symmetric rigid body with a constant external torque fixed with the axially symmetric body and perpendicular to its axis of symmetry, and with the initial absolute angular velocity also perpendicular to the symmetry axis and arbitrary initial body orientation.

**Theorem 4** *Given $p_0, q_0$ and $U$ real numbers, the solution of the dynamic problem of determining, at any time $t$, the absolute angular velocity of a rigid body having two equal principal moments of inertia about the principal body axes ($\sigma_1$) and ($\sigma_2$), initial angular velocity components $p(0) = p_0$, $q(0) = q_0$, and $r(0) = 0$ along the three principal axes, and subjected to a constant torque fixed with the axially symmetric body and perpendicular to its symmetry axis (without loss of generality we can assume that the torque is directed as the first axis of the coordinate system $B$), is the following*

$$^B\omega_{BN}(t) = \begin{Bmatrix} p_0 + U\,t \\ q_0 \\ 0 \end{Bmatrix}, \qquad (56)$$

*while the solution of the correspondent kinematic problem of determining, at any time $t$, the orientation of the body with respect to the inertial frame, i.e. of the coordinate system $B$ with respect to $N$, is given by*

$$R_{BN}(t) = R_{SN}(t), \qquad (57)$$

*where $R_{SN}(t)$ is obtained from Eq. 16 of Corollary 2, by substituting the subindex $B$ by $S$, the time-independent matrix $R_{BK}$ by the matrix $R_{SK}$ defined in Eq. 49, and the values of the initial conditions by $p'_0 = p_0, q'_0 = q_0, r'_0 = 0$.*

*Proof* In this case $\omega_e = 0$, i.e. the coordinate system $B$ is at rest with respect to the coordinate system $S$. In other words, in this case the axial symmetric body moves under the effect of the acting constant torque analogously to a spherical symmetric body. Therefore, the theorem statement follows directly from Eq. 3 and from Corollary 2.

5.3 Case without external torques

A solution to this classical case (Euler-Poinsot's spontaneous motion) is well known (see for instance Leimanis 1965). However, an original and simpler solution, as regards its derivation, is presented here below in terms of the rotation matrix. This solution is obtained by applying Hestenes' reduction theorem (Hestenes 1999). Hestenes 1999 reports a similar solution but he uses the not widely familiar notion of Spinor of rotation in order to describe the body orientation, while here below the solution is reported in terms of the rotation matrix.

The solution developed here for both the dynamics and kinematics are advantageous in that they are obtained with less algebraic steps than the classical solution, which relies on the geometrical construction of the body-fixed cone and the "space-fixed cone".

**Theorem 5** *Given $p_0, q_0, r_0$ real numbers, the solution of the dynamic problem of determining, at any time t, the absolute angular velocity of a rigid body having two equal principal moments of inertia about the principal body axes ($\sigma_1$) and ($\sigma_2$), initial angular velocity components $p(0) = p_0$, $q(0) = q_0$, and $r(0) = r_0$ along the three principal axes, and not subjected to external torques is the following*

$$^{B}\omega_{BN}(t) = \begin{Bmatrix} p_0 \cos(\omega_e\,t) + q_0 \sin(\omega_e\,t) \\ -p_0 \sin(\omega_e\,t) + q_0 \cos(\omega_e\,t) \\ r_0 \end{Bmatrix}, \qquad (58)$$

*with $\omega_e = \frac{I-I_3}{I} r_0$ (Eq. 34). Furthemore, the solution of the correspondent kinematic problem of determining, at any time t, the orientation of the body with respect to the inertial frame, i.e. of the coordinate system B with respect*

to $N$, is given by

$$R_{BN}(t) = R_{BS}(t)\, R_{SN}(t) \tag{59}$$

where $R_{SN}(t)$ is obtained from Eq. 6, with $\omega_{SN} = \omega_{SN}(0)$ as given by Eq. 32. Finally, $R_{BS}(t)$ is given by Eq. 27 with

$$f(t) = \left(\frac{I - I_3}{I}\right) r_0\, t. \tag{60}$$

*Proof* In this case, Eq. 25 states the conservation of the angular momentum

$$\underline{h}(t) = \underline{h}_0. \tag{61}$$

Therefore, the dynamic solution is straightforwardly given by following the development of Section 4, with the kinematic problem solved by considering Eq. 6. An alternative to the use of Eq. 6 is the use of the Euler's equation (see for instance Goldstein 1980)

$$R_{SN}(t) = \left\{\cos(\phi(t))\,\mathcal{I} + [1 - \cos(\phi(t))]\, a\, a^T - \sin(\phi(t))\, \Omega(a)\right\} R_{SN}(0), \tag{62}$$

where $\Omega(\bullet)$ is the skew-symmetric matrix function appearing also in Eq. 5, $\mathcal{I}$ indicates the three by three identity matrix, $a$ indicates the unit vector of the Euler's axis projected in either the $S$ or the $N$ coordinate system, and $\phi$ indicate the Euler's angle of rotation, which is satisfying, in this case, the following simple equation

$$\phi(t) = |{}^S\omega_{SN}(0)|\, t. \tag{63}$$

In writing Eq. 62 it has been taken into account that, in accordance with Assumptions 1, $R_{SN}(0) = R_{BN}(0)$. Finally, the axis of rotation $a$ can be immediately obtained by considering that it coincides with the direction of the angular velocity ${}^S\omega_{SN}(0)$, therefore yielding

$$a = \frac{{}^S\omega_{SN}(0)}{|{}^S\omega_{SN}(0)|}. \tag{64}$$

Interestingly, the overall spontaneous motion of an axially symmetric rigid body having three degrees of freedom is seen as intrinsically equivalent to the one degree of freedom motion of the "virtual" spherical body with respect to the inertial frame (motion of $S$ with respect to $N$). This motion captures the precession of the inertial symmetry axis of the physical body about the constant and inertially fixed absolute angular momentum vector. On the other hand, the motion of the coordinate system $B$ with respect to $S$ captures the relative spinning of the axially symmetric body about its axis.

## 6 Conclusions

For the first time, to the knowledge of the author, the exact analytic solutions, for both the kinematic and the dynamic problems, have been introduced in this paper for the rotational motion of a rigid body having revolution ellipsoid of inertia, in the following three cases

1. External torque constant in magnitude and parallel to the symmetry axis, with arbitrary initial angular velocity and body orientation.
2. External torque constant in magnitude, perpendicular to the symmetry axis and rotating at a constant rate about the symmetry axis, with arbitrary initial angular velocity and body orientation.
3. External torque and initial angular velocity perpendicular to the symmetry axis, with the torque constant in magnitude and fixed with the body and arbitrary initial body orientation.

The kinematic solutions are presented in terms of the rotation matrix.

In particular, the results of this paper are obtained by building upon the recently found exact analytic solution for the motion of a rigid body with a spherical ellipsoid of inertia, and by decomposing the motion of an axially symmetric rigid body into the combination of the motion of a "virtual" spherical body with respect to the inertial frame and that of the axially symmetric body with respect to this "virtual" body.

All of the analytical solutions introduced in this paper have been successfully checked against numerical solutions obtained through differential equations propagation, for sample cases.

Longuski 1991 and Tsiotras and Longuski 1991 give an exact analytic solution for the dynamic equation of an axially symmetric body which is applicable in a more general case (torque fixed with the body and having a generic direction) than the cases introduced in the present paper. Nevertheless the exact analytic solutions introduced in the present paper, when they apply, are critically advantageous because they are complete for both the dynamics and the kinematics.

In conclusion, the newly introduced exact analytic solutions are of high theoretical interest, as they add new elements to the set of few special cases for which a complete exact solution of the rotational motion of a rigid body exists. Additionally, they are also of interest from the applied mathematics point of view as they constitute significant new comparison cases for the validation and error analysis of approximate algorithms. In particular, the analytic solutions presented in this paper are valid for any length of time and rotation amplitude.

25